\documentclass[aps,prl,twocolumn,amsmath,amssymb,showpacs,floatfix,nofootinbib,superscriptaddress]{revtex4-1}

\usepackage{graphicx}
\usepackage{dcolumn}
\usepackage{xcolor}
\usepackage{bm}
\usepackage{natbib}
\usepackage{calligra}
\usepackage[T1]{fontenc}
\usepackage{egothic}
\usepackage[T1]{fontenc}
\newfont{\rsfsten}{rsfs10 scaled 1200}
\newfont{\rsfsseven}{rsfs10 scaled 1200}
\newfont{\rsfsfive}{rsfs10 scaled 1200}
\usepackage{epsfig}
\usepackage{units}
\usepackage[utf8]{inputenc}
\usepackage{multirow}

\advance \voffset by -0.5cm\relax

\def\lsim{\mathrel{\raise.3ex\hbox{$<$\kern-.75em\lower1ex\hbox{$\sim$}}}}
\def\gsim{\mathrel{\raise.3ex\hbox{$>$\kern-.75em\lower1ex\hbox{$\sim$}}}}

\begin{document}

\title{Bounds on Ultra-Light Hidden-Photon Dark Matter from 21cm at Cosmic Dawn}

\author{Ely D. Kovetz}
\affiliation{\vspace{0.05in}Department of Physics and Astronomy, Johns Hopkins University, Baltimore, MD 21218 USA}
\author{Ilias Cholis}
\affiliation{\vspace{0.05in}Department of Physics, Oakland University, Rochester, MI 48309 USA}
\affiliation{\vspace{0.05in}Department of Physics and Astronomy, Johns Hopkins University, Baltimore, MD 21218 USA}
\author{David E. Kaplan}
\affiliation{\vspace{0.05in}Department of Physics and Astronomy, Johns Hopkins University, Baltimore, MD 21218 USA}

\begin{abstract}
Ultra-light hidden-photon dark matter produces an oscillating electric field in the early Universe plasma, which in turn
induces an electric current in its ionized component whose dissipation results in heat transfer from the dark matter
to the plasma. This will affect the global 21cm signal from the Dark Ages and Cosmic Dawn. 
In this work we focus on the latter, in light of the reported detection by the EDGES collaboration of an absorption signal at frequencies corresponding to redshift $z\sim17$. By measuring the 21cm global signal, a limit can be placed on the amount of gas heating, and thus the kinetic mixing strength $\varepsilon$ between the hidden and ordinary photons can be constrained. 
Our inferred 21cm bounds on $\varepsilon$ in the mass range $10^{-23}\,{\rm eV}\lesssim m_\chi\lesssim10^{-13}\,{\rm eV}$ are the strongest to date. 
\end{abstract}


\maketitle


Dark matter remains an elusive ingredient of $\Lambda {\rm CDM}$, the concordance cosmological 
standard model. Despite various convincing measurements attesting to its cosmic abundance~\cite{Ade:2015xua,Freese:2017idy}, 
its makeup remains a puzzle. 
In recent decades, limits on models involving weakly-interacting 
massive particles (WIMPs)~\cite{Jungman:1995df} have been constantly tightening
\cite{Essig:2012yx,Aprile:2012nq,Akerib:2015rjg}, increasing the motivation to explore 
alternatives such as axions~\cite{Turner:1989vc,Raffelt:1990yz,Cadamuro:2011fd,Marsh:2015xka}, sterile neutrinos~\cite{Kusenko:2009up}, and various types of massive compact objects~\cite{1986ApJ...304....1P,Carr:1974nx,Bird:2016dcv,Clesse:2016vqa}.

Ultra-light hidden photons provide an appealing candidate for dark matter. These are (light) massive vector bosons 
that arise naturally in many theoretical setups, and which generically interact with the Standard Model (SM) 
through kinetic mixing with the ordinary photons. In principle, the allowed parameter space in coupling and mass is enormous, but the interaction
with SM photons opens up potential observational windows to probe them. 

One of the most promising such windows in the history of the Universe is the Cosmic Dawn era---the period where the first stars were born---which is accessible through the observation of the 21cm global signal.  
As this epoch marks the time where the average baryon temperature was at its lowest, it provides a unique opportunity to probe possible interactions
between baryons and dark matter, which under $\Lambda$CDM is much colder.

Excitingly, the Experiment to Detect the Global Epoch of Reionization Signature (EDGES)~\cite{Bowman:2018yin} recently reported a detection of an absorption profile centered at $78\,{\rm MHz}$ (corresponding to redshift $z\!\sim\!17$ if due to 21cm line emission from neutral hydrogen), with a best-fit amplitude more than twice the maximum allowed in $\Lambda$CDM. 
While explaining the depth of this profile is challenging, this measurement can be used as a test of various dark matter models which predict heating of the baryonic gas (which would reduce the absorption amplitude)~\cite{DAmico:2018sxd,Mitridate:2018iag}.

Ultra-light hidden-photon dark matter (HPDM) has  been shown to produce plasma heating as a 
result of its coupling to the SM electric current \cite{Dubovsky:2015cca}.
In this scenario, the HPDM has a mass $m_{\chi}$, while its coupling to the 
SM electric current is achieved via kinetic mixing between the HPDM and
the SM photon. Following the notation of Ref.~\cite{Dubovsky:2015cca}, the Lagrangian can be written as
\begin{eqnarray}
\mathcal{L} &=& -\frac{1}{4}F_{\mu\nu}F^{\mu\nu} -\frac{1}{4}\tilde{F}_{\mu\nu}\tilde{F}^{\mu\nu} +\frac{m^2}{2}\tilde{A}_{\mu}\tilde{A}^{\mu}  \cr
&&- {e\over (1+\varepsilon^2)^{1/2}}J^\mu\left(A_\mu + \varepsilon\tilde{A}_\mu\right),
\label{eq:Lagrangian}
\end{eqnarray} 	
where $A_\mu$ and  ${F}_{\mu\nu}$ ($\tilde{A}_\mu$ and $\tilde{F}_{\mu\nu}$) are the gauge field and field strength of the ordinary (hidden) photons,  
and $\varepsilon$ in the last term parameterizes the  kinetic mixing strength. 

The HPDM mass $m_{\chi}$ has to be 
compared to the effective SM photon mass in a given medium, which is set by its plasma 
frequency $\omega_{p}$. If $m_{\chi}$ is larger than $\omega_{p}$, then very 
efficient conversion of the HPDM into regular photons could have taken 
place in the early Universe, leading to either depletion of dark matter or a strong 
imprint on the cosmic microwave background (CMB) spectrum \cite{Arias:2012az}. 

In this \textit{Letter} we focus mainly on the alternative case where $m_{\chi} < \omega_{p}$.
The HPDM field on scales of the size of its de Broglie wavelength can cause an 
associated oscillating electric field which as a result will induce a SM electric current. 
That current will be damped as the intergalactic medium (IGM) has non-zero resistivity.
This mechanism can transfer energy from the dark matter to the ionized plasma,
heating it up, and in turn can affect the absorption of CMB photons in the hydrogen gas, which is observable 
through the 21cm brightness temperature contrast. 
While the ionized fraction of the  gas at cosmic dawn is low, we demonstrate that the effect on the 21cm 
signal is strong enough to place the strongest limits on this model to date in the ultra-light mass regime.

The IGM plasma frequency at Cosmic Dawn is \cite{RybickiLightman1986}
\begin{equation}
\omega_{p} = \left(\frac{4 \pi n_{e} \alpha}{m_{e}}\right)
 = 1.7 \times 10^{-14} \left(\frac{{n_{e}}}{2 \times 10^{-7} \, \textrm{cm}^{-3}}\right)^{1/2} \; \textrm{eV},
\label{eq:PlasmaFreq}
\end{equation}
where the number density $n_{e}$ of free electrons is strongly dependent on the
exact redshift, and the value $n_{e} = 2 \times 10^{-7} \textrm{cm}^{-3}$ corresponds 
to redshift $z\simeq17$, matching the central frequency $78\,{\rm MHz}$ of the anomalous absorption signal recently reported by EDGES \cite{Bowman:2018yin}. 
For redshifts in the range $z=13-20$, relevant for the expected era of Cosmic Dawn, 
the value of $n_{e}$ varies and can be up to a factor of two higher at $z = 20$ and a factor of 
two lower at $z = 13$. This change is mainly due to the average number 
density of hydrogen atoms changing by a similar factor (with the ionization 
fraction $x_{e}$ changing only by $\simeq 10 \%$ during that time). 
As we shall see, it is imperative to track the redshift dependence of $\omega_{p}$, 
as it can fall below the HPDM mass $m_{\chi}$ in the redshift range between baryon-photon 
decoupling and Cosmic Dawn, thereby abruptly weakening the baryon heating effect.

To calculate the heating rate of the plasma due to the HPDM field, 
we follow the treatment of Ref.~\cite{Dubovsky:2015cca}.
The induced motion of electrons and ions in the plasma results in collisional friction.
The collision frequency is given by
\begin{equation}
\nu = \frac{4 \sqrt{2 \pi} \alpha^{2} n_{e}}{3 m_{e}^{1/2}T_{e}^{3/2}} \, 
\textrm{ln} \left(\frac{4\pi T_{e}^{3}}{\alpha^{3}n_{e}} \right)^{1/2}.  
\label{eq:ElecIonCollFreq}
\end{equation}
Taking as a reference temperature for  $T_{e}$ that of the baryons one gets 
that the electron-ion collision frequency is reduced for decreasing redshifts from
$\simeq 2.5 \times 10^{-22}$ eV at $z=20$ to $\simeq 2 \times 10^{-22}$ at 
$z=17$, and to  $\simeq 7 \times 10^{-24}$ at $z=13$. This means that the heat transfer
due to the friction term associated with these collisions is faster at earlier stages (assuming reionization of the gas by radiation from the first stars and galaxies can be neglected). 

The dissipation of the induced oscillation in the plasma  is described by $\gamma_{\chi}$, the imaginary part of the 
oscillation frequency $\omega \equiv \omega_{\chi} + i \,\gamma_{\chi}$.  For 
$\varepsilon \ll 1$, $\omega_{\chi} = m_{\chi}$ and~\cite{Dubovsky:2015cca}
\begin{eqnarray}
     \gamma_{\chi} = \begin{cases} - \nu \frac{m_{\chi}^{2}}{2 \omega_{p}^{2}} \frac{\varepsilon^{2}}{1+\varepsilon^{2}}
      \,\,\, \textrm{for} \,\,\,  m_{\chi} \ll \omega_{p}, \\  
      - \nu  \frac{\varepsilon^{2}}{1+\varepsilon^{2}} \frac{\omega_{p}^{2}}{2 m_{\chi}^{2}}  \,\,\,  \textrm{for}  \,\,\, 
      m_{\chi} \gg \omega_{p} .
     \end{cases}
\label{eq:gammah}
\end{eqnarray}
In addition to the $m_{\chi} \ll \omega_{p}$ case, we include here also the opposite scenario 
of $m_{\chi} \gg \omega_{p}$. While Eq.~\eqref{eq:gammah} is strictly valid only at these  
extremes, we will naively connect the two regimes to study the full $(m_{\chi}, \varepsilon)$
parameter space. 

The resulting heat transfer rate $\dot Q_b$ from the HPDM field to the early Universe plasma is then given by
\begin{equation}
\dot Q_b=2 |\gamma_{\chi}| \rho_{\chi},
\label{eq:DMcooling}
\end{equation}
where $\rho_{\chi}$ is the energy density of the HPDM. 

In order to calculate the effect of this heating term, Eq.~\eqref{eq:DMcooling}, on the 21cm brightness temperature, 
it must first be properly incorporated in the evolution of the baryon gas temperature, which includes other heating and cooling mechanisms~\cite{Munoz:2015bca}. 
We describe this calculation next.

The 21cm brightness temperature contrast with respect to the CMB temperature $T_{\rm CMB}$ is given by \cite{Loeb:2003ya,Pritchard:2011xb,Kovetz:2018zan}
\begin{equation}
T_{\rm 21}(z)= \frac{T_s-T_{\rm CMB}}{1+z}\left(1-e^{-\tau}\right),~~\tau=\frac{3T_{*}A_{10}\lambda_{21}^3  n_{\rm HI}}{32\pi T_s H(z)}, 
\label{eq:T21}
\end{equation}
where $\tau$ is the optical depth for the hyperfine transition, $T_{*}=0.068\,{\rm K}$ is the energy difference between the two hyperfine levels, $A_{10}$ is the Einstein-A coefficient of the  transition, $\lambda_{21}\approx21.1\,{\rm cm}$ is the emission wavelength, and $n_{\rm HI}$ is the neutral hydrogen density.
The absorption (or emission) amplitude depends on $T_s$, the spin temperature of the gas, which parameterizes the ratio between the populations of the hyperfine triplet and singlet states.

As the processes that determine the spin temperature involve large astrophysical uncertainties, there is no exact prediction for the signal, Eq.~\eqref{eq:T21}, during Cosmic Dawn. Different astrophysical models yield values in the $z=13-20$ redshift range that differ by more than an order of magnitude~\cite{Cohen:2016jbh}.
However, under $\Lambda$CDM, there is an absolute minimum value for $T_{\rm 21}$, obtained when the spin temperature equals the gas temperature. Setting $T_s= T_b$ in Eq.~\eqref{eq:T21} one gets roughly $T^{\rm min}_{\rm 21}(z\sim17)\simeq-207\,{\rm mK}$.

To take into account the heating of the gas due to the HPDM, we need to evolve the baryon temperature $T_b$, starting from when the baryons are effectively coupled to the CMB via Compton scattering, and including the new heating term, Eq.~\eqref{eq:DMcooling}. The equation for $T_b$ is therefore
\begin{equation}
\dfrac{d T_b}{da } = -2\dfrac{T_b}{a} + \dfrac {\Gamma_C}{a H} (T_{\rm CMB}-T_b) + \dfrac {2 \dot Q_b}{3 a H n_H (1+f_{\rm He}+x_e)}, 
\label{eq:Tgas}
\end{equation}
where $H$ is the Hubble parameter and $\Gamma_C$ is the Compton interaction rate,
which depends on the free-electron density $n_e$ and on $x_e = n_e/n_H$, the free-electron fraction. We solve for $x_e(a,T_b)$ in tandem with Eq.~\eqref{eq:Tgas} \cite{Munoz:2015bca,Kovetz:2018zan}. 
We note that for $\varepsilon \ll 1$, as we study here, the third term in Eq.~\eqref{eq:Tgas} is $\sim10^{20}\; \varepsilon^2\, {\rm K}$ for $m_{\chi} = \omega_{p}$ at 
redshift $z=17$. This already gives us a rough estimate that to avoid any heating 
$\varepsilon$ should not exceed $\mathcal{O}(10^{-10})$ at $m_{\chi} \sim 10^{-14}$ eV. 
  
\begin{figure}[b!]
\centering
\includegraphics[width=\columnwidth]{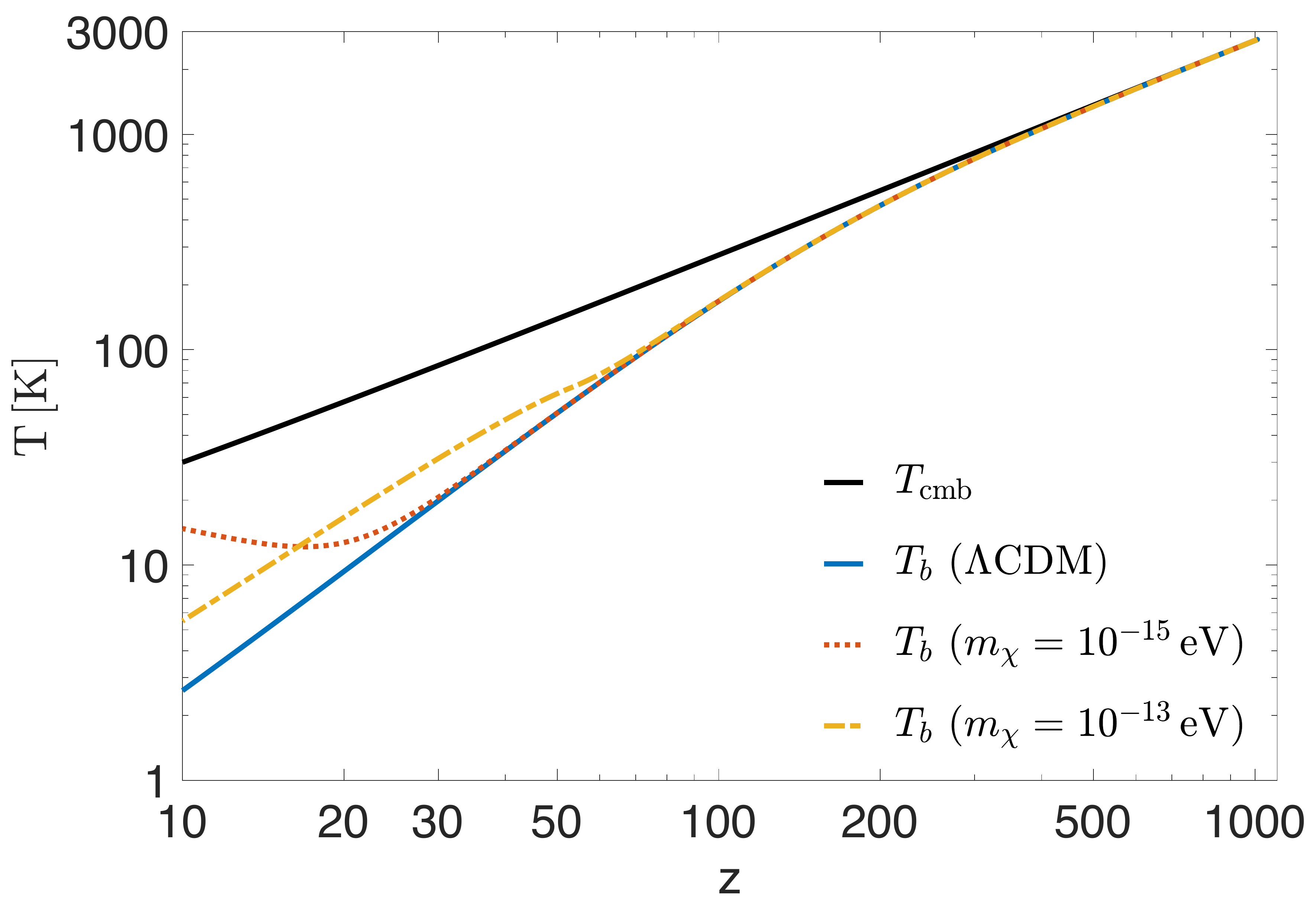}
\caption{The evolution of the baryon temperature, with and without heating due to HPDM, for different values of $m_\chi$, setting $\varepsilon$ in each case to yield a 21cm brightness temperature $T_{\rm 21}(z\!=\!17)\!=\!-100\,{\rm mK}$. For masses $m_\chi\gtrsim1.7\times10^{-14}\,{\rm eV}$, the heating weakens once the plasma frequency $\omega_{p}$ falls below the mass (as discussed in the text, for $m_\chi=1.7\times10^{-14}\,{\rm eV}$ this transition happens precisely at redshift $z=17$, see Eq.~\eqref{eq:PlasmaFreq}).}
\label{fig:BaryonTemperature}
\end{figure}

The evolution of the baryon temperature in the presence of heating by HPDM is shown in Fig.~\ref{fig:BaryonTemperature}. 
We show in solid lines the temperature evolution of the CMB and of the baryons under $\Lambda$CDM, from recombination to the end of Cosmic Dawn. The baryon temperature at $z=17$ is roughly $T_b\sim 8\,{\rm K}$. Dashed lines show the heating effect on the baryons, for HPDM masses in the two regimes described above, with $\varepsilon$ chosen to yield $T_b\sim 10\,{\rm K}$. 

We see that for low masses, $m_\chi\ll10^{-14}\,{\rm eV}$, the heating is monotonous, increasing with time. If the kinetic mixing parameter is large enough, this can lead to non-negligible heating of the gas. In the other regime, $m_\chi\gg10^{-14}\,{\rm eV}$, the heating is effective so long as the plasma frequency remains larger than the HPDM mass. Once it falls below it (which happens at different redshifts for different masses), the heating significantly weakens and the baryons again cool due to the Hubble expansion.

\begin{figure}[h!]
\centering
\includegraphics[width=\columnwidth]{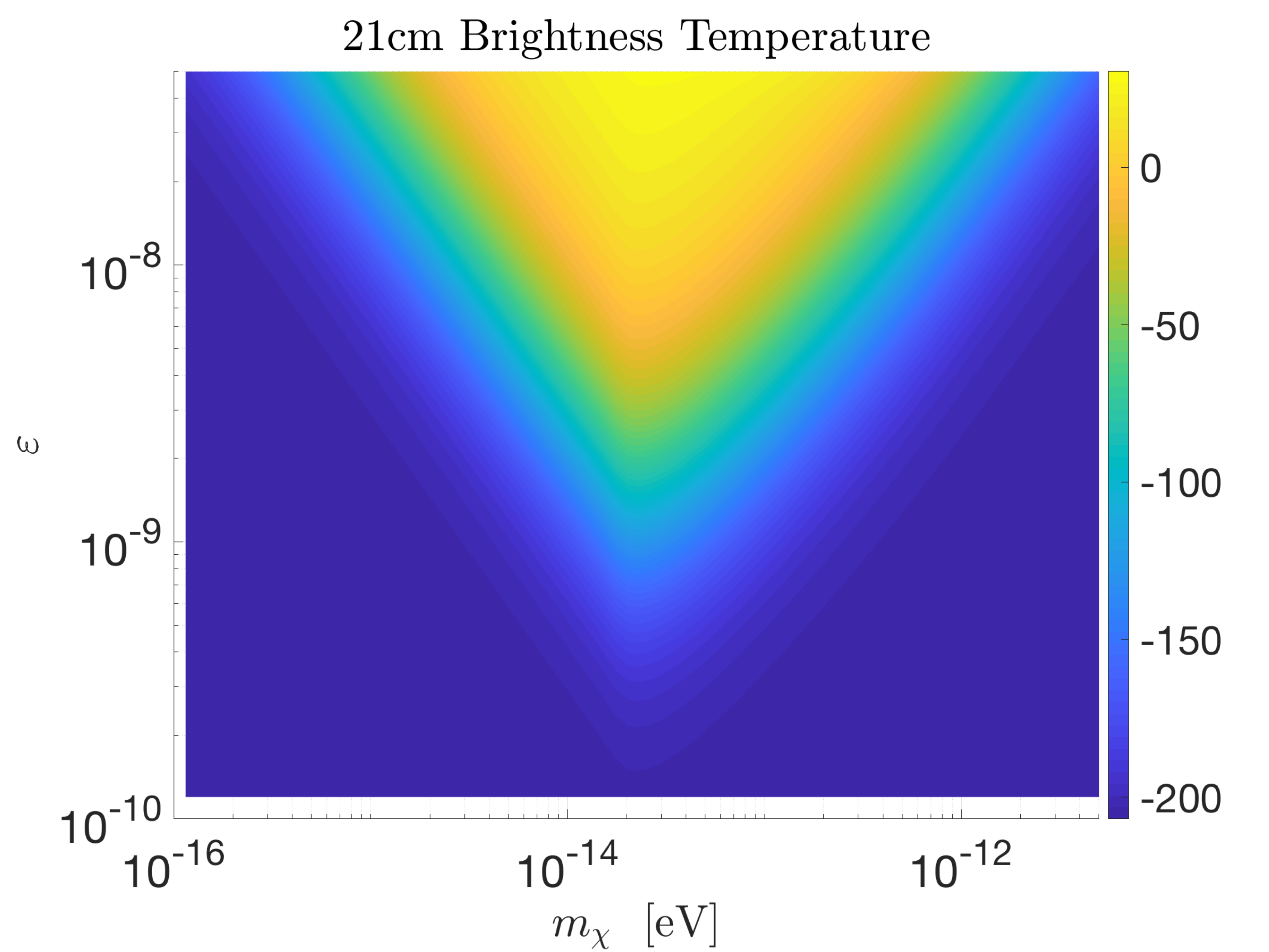}
\caption{The 21cm brightness temperature as a function of the kinetic mixing parameter $\varepsilon$ and the HPDM mass $m_\chi$. This map exhibits clear gradients from $T_{\rm 21}\sim-200\,{\rm mK}$ to $\sim\!30\,{\rm mK}$.}
\label{fig:21cmBrightnessTemperature}
\end{figure}

Combining Eqs.~\eqref{eq:DMcooling}, \eqref{eq:T21} and \eqref{eq:Tgas}, we show in Fig.~\ref{fig:21cmBrightnessTemperature} a map of the 21cm brightness temperature in the $(\varepsilon,m_\chi)$ parameter space. 
The behavior with respect to the HPDM mass can be understood from Eqs.~\eqref{eq:gammah} and \eqref{eq:DMcooling}. On the two sides of $m_\chi\sim10^{-14}\,{\rm eV}$, the slopes of the contours are approximately opposite. For that mass a kinetic mixing strength of $\varepsilon\sim10^{-8}$ would be enough to erase any absorption signal at Cosmic dawn (if one increases $\varepsilon$, at some point the signal will be in emission).

These results can be used to derive constraints on $\varepsilon$, in light of the strong absorption signal of EDGES~\cite{Bowman:2018yin}.
We note, though, that its amplitude, $T_{\rm 21}(z\sim17)=-500^{+200}_{-500}\,{\rm mK}$ (at $99\%$ confidence), lies well below (roughly $3.8\sigma$) the expectation from $\Lambda$CDM. If confirmed, this makes the placing of bounds on any gas heating mechanism nonstraightforward, as it may require some cooling which would also have to be compensated for\footnote{The EDGES signal could be explained e.g.~by cooling as a result of baryon--DM scattering~\cite{Tashiro:2014tsa,Munoz:2015bca,Xu:2018efh,Barkana:2018lgd,Munoz:2018pzp,Fialkov:2018xre,Berlin:2018sjs,Barkana:2018qrx,Slatyer:2018aqg,Kovetz:2018zan,Boddy:2018wzy}, or alternatively by new sources of radio emission~\cite{Feng:2018rje,Fraser:2018acy, Ewall-Wice:2018bzf,Pospelov:2018kdh}; earlier kinetic decoupling of baryons from CMB photons \cite{Hill:2018lfx,Falkowski:2018qdj,Poulin:2018dzj}; or foreground residuals~\cite{Hills:2018vyr}.}.  
We will follow here the simple approach of Refs.~\cite{DAmico:2018sxd,Mitridate:2018iag}, and set our bounds to correspond to heating strengths that would yield a 21cm brightness temperature of $T_{\rm 21}=-100\,{\rm mK}$ (or $-50\,{\rm mK}$) in the limit of infinite Lyman-$\alpha$ coupling, which as explained above, would otherwise (under $\Lambda$CDM) yield an amplitude  $T_{\rm 21}\sim-200\,{\rm mK}$.

\begin{figure}[t!]
\centering
\includegraphics[width=\columnwidth]{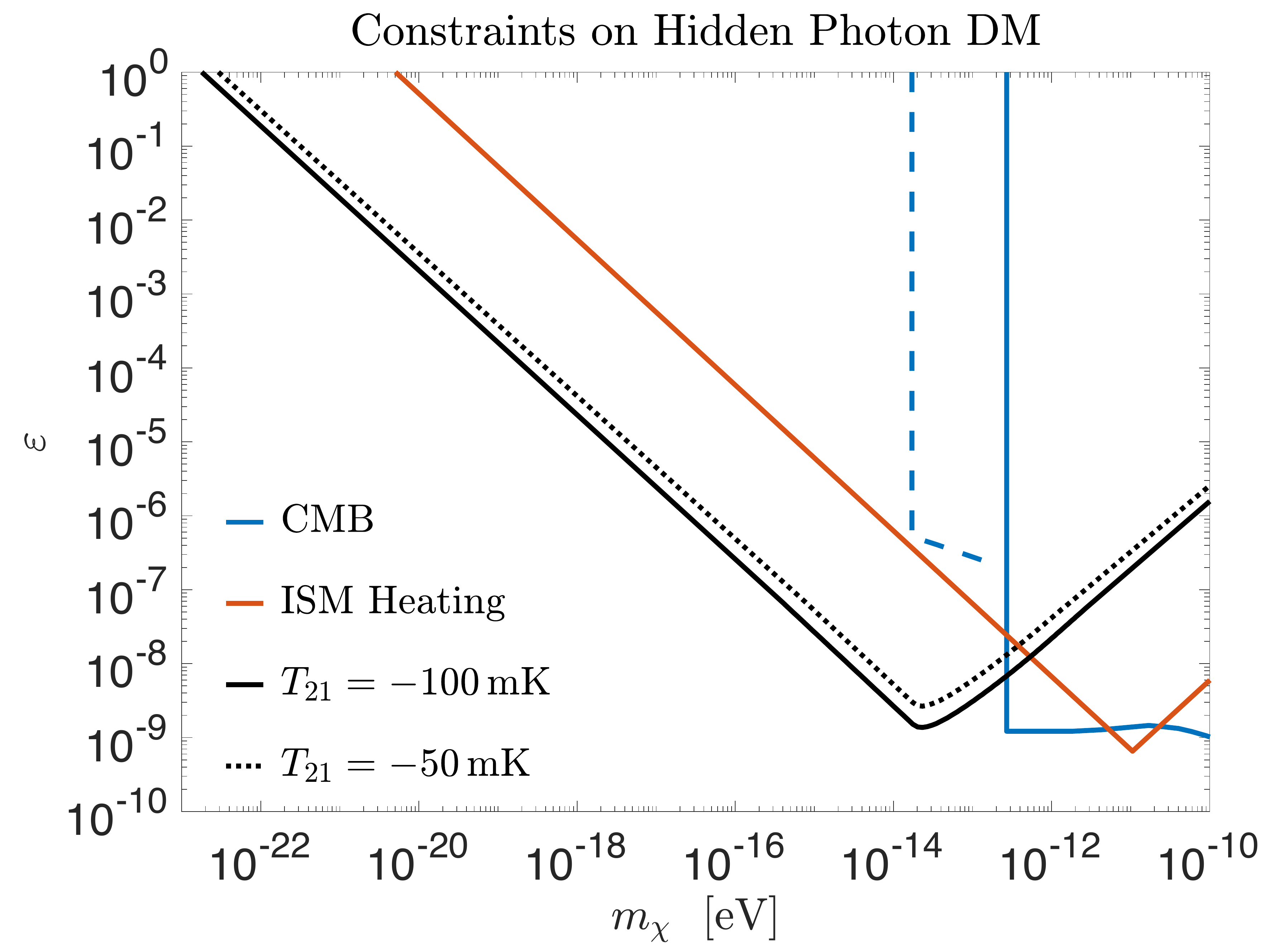}
\caption{Predominant bounds on the kinetic mixing parameter $\varepsilon$ for different HPDM masses $m_\chi$.  We show constraints from ISM heating~\cite{Dubovsky:2015cca} ({\it red}) and from the CMB~\cite{Arias:2012az,Chaudhuri:2014dla} ({\it blue}). Our inferred 21cm bounds from requiring that $T_{21}=-100\,{\rm mK}$ or $T_{21}=-50\,{\rm mK}$ ({\it black} and {\it dashed-black}) are  two orders of magnitude stronger for $m_\chi\lesssim10^{-14}\,{\rm eV}$ and the only ones to penetrate the fuzzy-DM mass range  $10^{-23}\,{\rm eV}\!\lesssim m_\chi\lesssim10^{-20}\,{\rm eV}$.}
\label{fig:21cmConstraints}
\end{figure}

This requirement leads to the limits shown in Fig.~\ref{fig:21cmConstraints}. The solid (dashed) black lines correspond to the limits inferred from requiring $T_{21}=-100\,{\rm mK}$ ($T_{21}=-50\,{\rm mK}$). The figure also shows competing limits from two different sources: observed interstellar medium (ISM) gas clouds in the Milky Way (MW), and the CMB. The heating of gas clouds in the MW follows the same principles as the IGM heating we have described here. Ref.~\cite{Dubovsky:2015cca} calculated this heating rate (at redshift $z\!=\!0$) and by requiring it to be smaller than the observed cooling rate in the MW ISM, derived strong constraints on $\varepsilon$ that are competitive with the CMB for masses $m_\chi\lesssim10^{-11}\,{\rm eV}$ and extend well into the ultra-light regime, all the way to $m_\chi\gtrsim10^{-20}\,{\rm eV}$, see Fig.~\ref{fig:21cmConstraints}. It would be interesting to compare these with constraints from IGM heating at higher redshift, based on the Lyman-$\alpha$ forest (see Ref.~\cite{Munoz:2017qpy} for a similar analysis).

Meanwhile, the are several effects that HPDM could have on the CMB if its mass is higher than the effective mass of the ordinary photons, which is set by the plasma frequency. First, if resonant conversion of hidden photons to ordinary photons occurs at high redshift, before recombination, this would change the relative number of neutrinos and baryons relative to photons, which in turn would decrease $N_{\rm eff}$, the number of effective neutrino degrees of freedom. This quantity is well constrained by Planck~\cite{Aghanim:2018eyx} (and one could moreover impose consistency between the Big-Bang nucleosynthesis and CMB measurements~\cite{Cooke:2017cwo,Zavarygin:2018dbk}). A second observable effect is spectral distortions caused by the photon injection, in the form of chemical potential $\mu$ distortion or Compton-$y$ distortion, depending on the timing of the HPDM energy dump. These two effects, however, are only efficient in constraining HPDM with mass $\gtrsim10^{-10}\,{\rm eV}$, which is still larger than the plasma frequency at redshifts $z\gtrsim1100$. 

The most dominant bound on the CMB in our mass range of interest comes from the simple requirement that the total depletion of hidden photons from recombination to the present day does not amount to a change in the dark matter energy density that would violate the agreement between the CMB power spectrum constraints and measurements of the average local DM density. 
This HPDM limit~\cite{Arias:2012az,Chaudhuri:2014dla} is shown in Fig.~\ref{fig:21cmConstraints} (solid blue line). 
CMB spectral distortion bounds restricting thermal production of hidden photons at the resonance extend to slightly lower masses~\cite{Mirizzi:2009iz} (dashed blue), but  are weaker than the ISM limits.
Masses lower than $\sim\!10^{-14}\,{\rm eV}$ are smaller than the plasma frequency in the IGM today, rendering constraints from resonant conversion ineffective\footnote{The gas heating we consider in the smaller HPDM mass regime can also affect the CMB power spectrum through the change it induces in the ionization fraction~\cite{Chen:2003gz,Poulin:2016anj}, but this effect is small.}. 
  
In conclusion, our derived bounds on ultra-light HPDM are the most stringent across roughly ten orders of magnitude, $10^{-23}\,{\rm eV}\lesssim m_\chi\lesssim10^{-13}\,{\rm eV}$. They are stronger than the competing limits by more than two orders of magnitude in the $m_\chi\lesssim10^{-13}\,{\rm eV}$ mass range. Our bounds for $10^{-23}\,{\rm eV}\lesssim m_\chi\lesssim10^{-20}\,{\rm eV}$ now place non-trivial limits on the vector version of fuzzy DM~\cite{Hu:2000ke,Hui:2016ltb}.

We note that astrophysical effects can play a role in setting the actual 21cm absorption amplitude, and disentangling them from the possible influence of dark matter may not necessarily be trivial. As shown in Refs.~\cite{Kovetz:2018zan,Venumadhav:2018uwn}, neglecting additional sources of heating one can easily underestimate by factors of a few the minimum amplitude of DM--baryon scattering that could explain the anomalous EDGES signal. A more conservative approach to accommodate that would be to take $T_{21}\!=\!0\,{\rm mK}$ when setting our bounds in Fig.~\ref{fig:21cmConstraints}. Yet even in that case, the resulting bounds would weaken by less than an order of magnitude (see Fig.~\ref{fig:21cmBrightnessTemperature} for $T_{21}=0\,{\rm mK}$), and still remain significantly tighter than those in existing literature.

The ultimate 21cm probe of models such as HPDM will be the 21cm power spectrum~\cite{Cohen:2017xpx}, which can provide additional constraining power to the global signal, and also be used to distinguish between different sources of heating based on their spectral contribution~\cite{Fialkov:2018xre,Kaurov:2018kez}\footnote{See Refs.~\cite{Munoz:2015bca,Munoz:2018jwq} for the case of DM--baryon scattering.}. Fortunately, many experiments are in pursuit~\cite{Kovetz:2017agg}. The Cosmic Dawn 21cm signal has yet to lend its final word.

\acknowledgements
We thank Kimberly Boddy, Julian Mu\~noz and Vivian Poulin for useful discussions. 
This work was supported by NSF Grant No. 0244990, NASA  NNX17AK38G and the Simons Foundation.

\end{document}